\documentclass[a4paper,12pt]{article}
\usepackage{pdflscape}
\usepackage{amsmath}
\usepackage{cite}
\usepackage{amssymb}
\usepackage{dcolumn}
\usepackage{bm}
\usepackage{color}
\usepackage{epsfig}
\usepackage{amsfonts}
\usepackage{graphicx}

\usepackage{dcolumn}
\usepackage{indentfirst}
\usepackage{authblk}
\usepackage{slashed}
\usepackage{appendix}

\usepackage{booktabs}
\AtBeginDocument{
\heavyrulewidth=.08em
\lightrulewidth=.05em
\cmidrulewidth=.03em
\belowrulesep=.65ex
\belowbottomsep=0pt
\aboverulesep=.4ex
\abovetopsep=0pt
\cmidrulesep=\doublerulesep
\cmidrulekern=.5em
\defaultaddspace=.5em
}

\usepackage[colorlinks=false]{hyperref}
\hypersetup{citecolor=Blue}

\begin{document}

\vspace{80pt}

\centerline{\LARGE  Holographic Schwinger Effect In a Step Dilaton Background}
\vspace{40pt}
\centerline{
 Sara Tahery$^{\ast}$ and Qin Chang$^{\dagger}$
\let\thefootnote\relax\footnote{$^{\ast}$saratahery@htu.edu.cn}
 
\let\thefootnote\relax\footnote{$^{\dagger}$  changqin@htu.edu.cn} 
}
\vspace{30pt}
{\centerline {$^{\ast,\dagger}${\it Institute of Particle and Nuclear physics,  
Henan Normal University,  Xinxiang 453007, China
}}

\vspace{40pt}
\begin{abstract}

We investigate the holographic Schwinger effect in a confining background with a step dilaton profile, which induces a sharp transition between ultraviolet and infrared regimes and provides a qualitatively distinct realization of confinement. Within this framework, the quark--antiquark potential is extracted from the classical configuration of a fundamental string, allowing for a direct analysis of vacuum instability and pair production. In the absence of a magnetic field, the step dilaton leads to a significantly sharper suppression of the potential barrier as the electric field increases, implying an enhanced sensitivity of the critical electric field compared to smooth soft-wall models and demonstrating that the abrupt geometric transition qualitatively enhances the onset of vacuum decay. Incorporating an external magnetic field through the Dirac--Born--Infeld action, we find a nontrivial and amplified deformation of the potential barrier, resulting in a pronounced shift of the critical electric field that depends on both the magnitude and orientation of the magnetic field. Overall, the step dilaton background exhibits a substantially stronger response of the Schwinger effect to external electromagnetic fields than conventional soft-wall models, providing a novel mechanism for controlling pair production and highlighting the crucial role of dilaton structure in non-perturbative dynamics of holographic QCD.

\end{abstract}

\newpage

\tableofcontents

\section{Introduction}
The gauge/gravity duality, originally proposed in \cite{adscft}, has provided a powerful framework for studying strongly coupled quantum field theories via their dual gravitational descriptions. Within this approach, many non-perturbative aspects of quantum chromodynamics (QCD), such as confinement and vacuum structure, can be investigated holographically. In particular, bottom-up AdS/QCD models incorporating nontrivial dilaton profiles have been widely employed to reproduce essential features of hadronic physics. The soft-wall model introduced in \cite{Karch2006} successfully realizes linear confinement through a smooth dilaton background, while subsequent developments, including improved dilaton constructions \cite{Cox2015} and dynamical holographic QCD models \cite{Li2013}, have refined the description of infrared physics and chiral symmetry breaking.
 
One of the most prominent non-perturbative phenomena in quantum field theory is the Schwinger effect, namely the production of particle--antiparticle pairs from the vacuum under a strong external electric field. Its holographic realization was established in \cite{Semenoff2011}, where the pair creation process is mapped to the dynamics of a fundamental string in the bulk geometry. In this setup, the quark--antiquark potential is obtained from the classical string configuration, providing a geometric description of the tunneling process. This framework has been further developed in various directions: the potential analysis of the holographic Schwinger effect was systematically studied in \cite{Sato2013,Sato1309}, while extensions to finite temperature and chemical potential were investigated in \cite{Hou2018,Zhang2016}. Additional generalizations include non-relativistic backgrounds \cite{Kazem1504}, higher-derivative corrections such as Gauss--Bonnet gravity \cite{Zi-qiang2018}, anisotropic and flavor-dependent systems \cite{Lin2025,Wen2506}, translational symmetry breaking systems \cite{sara2025} and dynamical AdS/QCD settings \cite{zhu2021}, all of which have contributed to a deeper understanding of vacuum instability in strongly coupled systems.
 
The influence of external electromagnetic fields on the Schwinger effect has also attracted considerable attention. In particular, the combined effects of electric and magnetic fields were incorporated holographically in \cite{satoyo1303}, where the modification of the critical electric field was analyzed. Further insights into the behavior of critical fields and vacuum instability were provided in \cite{Bolo1210}, while the connection to the Euler--Heisenberg effective action and magnetic instabilities was explored in \cite{Koji1403}. More recently, the impact of magnetic fields on the potential barrier and pair production process has been examined in magnetized backgrounds \cite{zhou1912}. These studies demonstrate that external fields can significantly alter the structure of the potential and the corresponding pair creation rate.
 
Despite these extensive developments, most existing analyses have been restricted to pure AdS or smooth soft-wall geometries. In contrast, alternative dilaton profiles can lead to qualitatively different realizations of confinement. In particular, step-like dilaton configurations introduce a sharp transition between ultraviolet and infrared regimes, effectively modeling an abrupt change in the coupling structure of the dual field theory. Such backgrounds differ fundamentally from the smooth behavior of soft-wall models and are expected to induce nontrivial modifications in the holographic description of non-perturbative processes. However, a systematic investigation of the holographic Schwinger effect in the presence of a step dilaton profile is still lacking in the literature, which motivates the present study.
 
In this work, we address this gap by analyzing the holographic Schwinger effect in a background characterized by a step dilaton field. We first construct the corresponding holographic setup and clarify its physical interpretation in terms of confinement and energy scales. We then evaluate the quark--antiquark potential using the holographic method, both in the absence and in the presence of an external magnetic field, thereby providing a unified framework to examine the role of electromagnetic fields in this geometry. In addition, we compare our analysis with existing results obtained in soft-wall models \cite{Ding2020}, in order to highlight the role of the dilaton structure in shaping the Schwinger effect. The organization of the paper is as follows. In section \ref{sec:model}, we introduce the holographic background with the step dilaton field and discuss its main physical features. In section \ref{sec:Dilaton}, we present the analysis of the Schwinger effect for both vanishing and finite magnetic fields. Finally, in section \ref{sec:summary} we summarize our results and outline possible future directions.
\section{Holographic Setup with Step Dilaton Background} \label{sec:model}

The background geometry is taken to be the AdS--Schwarzschild metric,
\begin{equation}\label{metric}
ds^2= \frac{r^2}{L^2} \Big( -f(r)\,dt^2+dx_i^2\Big)+\frac{L^2}{r^2\,f(r)}dr^2,
\end{equation}
where $f(r)=1-\frac{r_h^4}{r^4}$, with $r_h$ denoting the black hole horizon and $L$ the AdS radius. This geometry provides a finite-temperature background for the dual gauge theory, with the Hawking temperature corresponding to the temperature of the boundary system. The presence of the horizon introduces a natural infrared scale, which plays an important role in nonperturbative phenomena.

The dynamics of a fundamental string probing this geometry is governed by the Nambu--Goto action. In the presence of a nontrivial dilaton field, the action is modified as
\begin{equation}\label{NGaction}
S_{NG}=\frac{1}{2\pi \alpha'}\int d\tau\,d\sigma\,e^{\phi(r)/2}\,\sqrt{-\det \mathcal{G}_{ab}},
\end{equation}
where $\mathcal{G}_{ab}$ is the induced metric on the string worldsheet. The exponential dilaton factor arises naturally in the string-frame formulation and reflects the direct coupling between the fundamental string and the dilaton background. Physically, this coupling induces a position-dependent effective string tension, thereby modifying the energetics of string configurations and directly affecting observables such as interquark potentials and pair production rates.

\subsection{Physical Interpretation of Dilaton Profile}

In holographic QCD models, particularly within the soft-wall AdS/QCD framework, the dilaton field $\phi(r)$ plays a central role in breaking conformal symmetry and implementing confinement in the infrared regime. The dilaton profile $\phi(r)$ is introduced as a background scalar field depending on the holographic radial coordinate $r$, which is dual to the energy scale of the boundary gauge theory. Consequently, large values of $r$ correspond to the ultraviolet (UV) regime, while smaller values of $r$ probe the infrared (IR) dynamics.

A phenomenologically motivated parametrization of the dilaton profile is given by
\begin{equation}
\phi(r) = A + 1 - A \tanh\big((r - \lambda) \kappa\big),
\label{eq:dilaton}
\end{equation}
which provides a smooth but controllable interpolation between two asymptotic regimes. In the UV limit ($r \to \infty$), the dilaton approaches $\phi \to 1$, corresponding to an approximately conformal phase of the dual theory. In contrast, in the IR region, the dilaton deviates from this value, encoding nonperturbative effects such as confinement and condensate formation. The hyperbolic tangent structure ensures a continuous crossover between these regimes, effectively mimicking the renormalization group flow from a UV fixed point toward an IR-dominated phase.

The parameters appearing in Eq.~\eqref{eq:dilaton} have clear and distinct physical interpretations. The parameter $A$ controls the amplitude of the dilaton deformation, determining the magnitude of the variation between the UV and IR regimes. Increasing $A$ enhances the strength of the infrared modification and can be associated with stronger nonperturbative effects in the dual field theory. The parameter $\lambda$ determines the radial position of the transition, thereby setting the characteristic energy scale at which the system departs from conformality. Shifting $\lambda$ effectively moves the location of this crossover along the holographic direction. The parameter $\kappa$ governs the sharpness of the transition: small values of $\kappa$ correspond to smooth, gradual crossovers, while large values produce a sharper, more localized transition, approaching a step-like profile. The constant term ensures a properly normalized UV limit, providing a consistent reference point for comparing different dilaton configurations.

This parametrization allows for a systematic exploration of how localized versus smooth infrared deformations affect physical observables. In particular, it enables a direct comparison between conventional soft-wall--type profiles and step-like dilaton configurations, highlighting the role of scale localization and abrupt transitions in strongly coupled systems.
\begin{figure}[htbp]
\centering
\includegraphics[width=8cm]{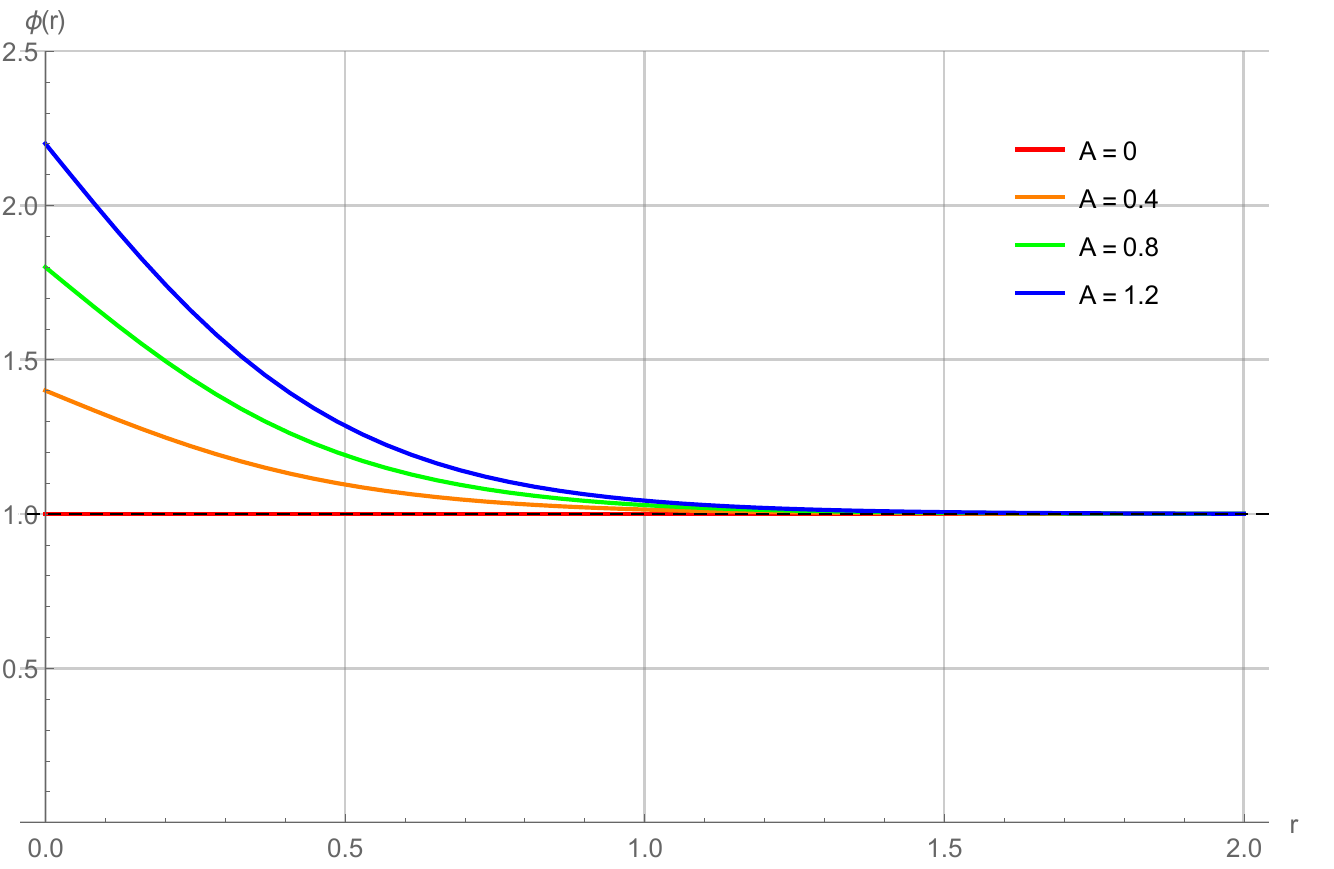}
\includegraphics[width=8cm]{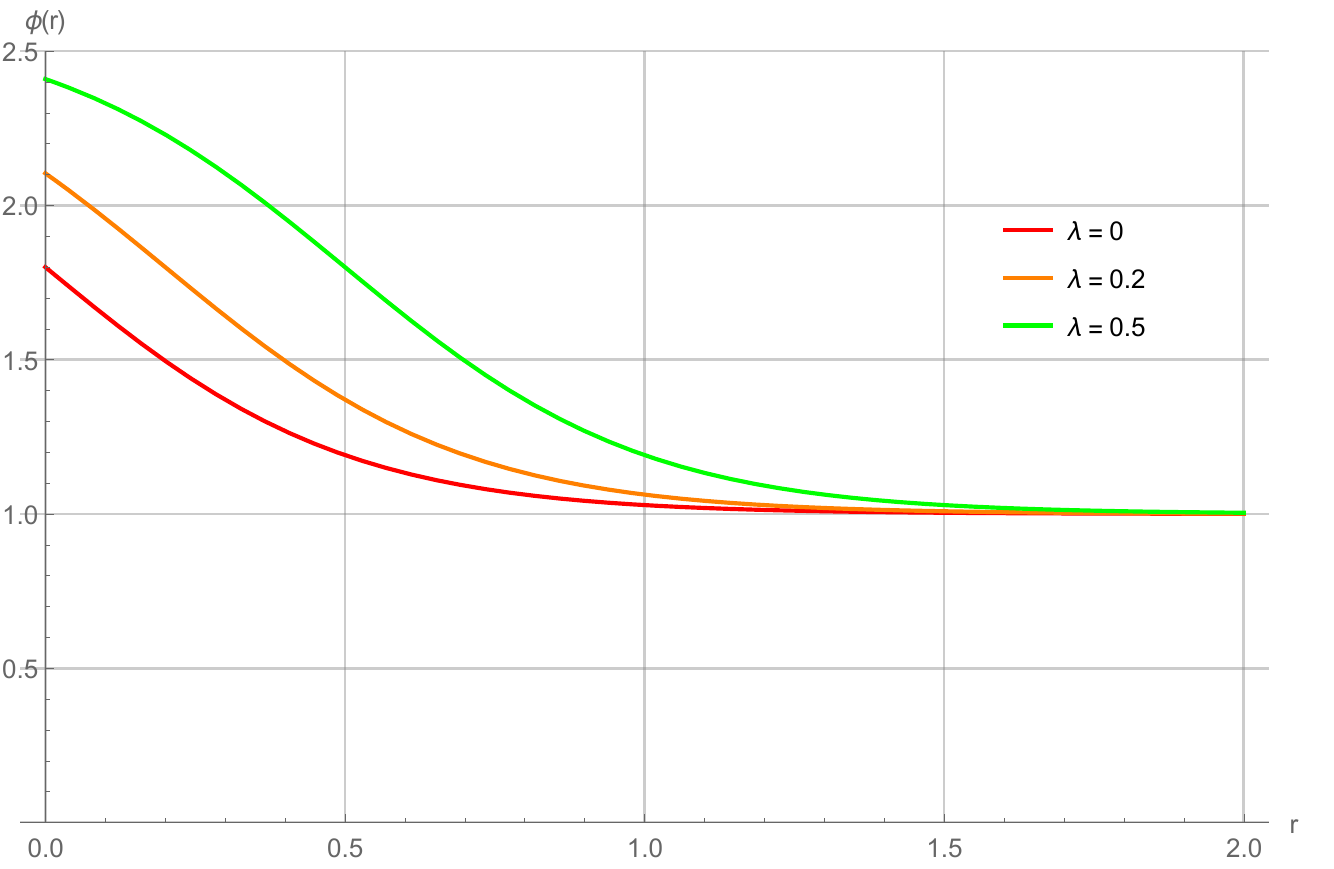}
\includegraphics[width=8cm]{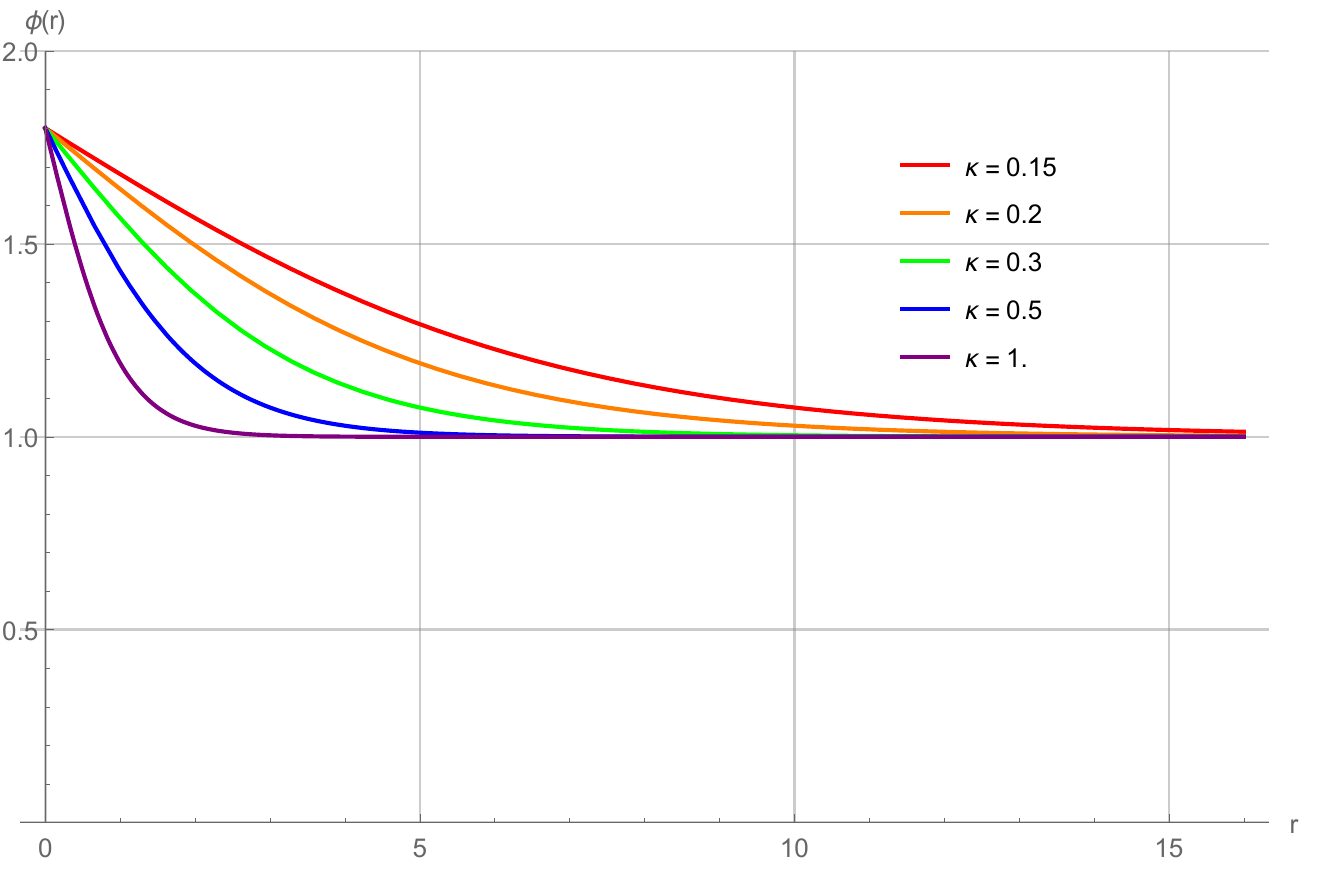}
\caption{The dilaton profile $\phi(r)=A+1-A\tanh\big((r-\lambda)\kappa\big)$ for different values of the parameters $A$, $\lambda$, and $\kappa$.}
\label{fig:path_independence}
\end{figure}

\textbf{Fig.~\ref{fig:path_independence}} shows the dilaton profile $\phi(r)=A+1-A\tanh\!\big((r-\lambda)\kappa\big)$ for representative variations of the parameters $A$, $\lambda$, and $\kappa$. The parameter $A$ controls the amplitude of the step, determining the magnitude of the variation between the ultraviolet (UV) and infrared (IR) regions. As $A$ increases, the separation between the asymptotic values of the dilaton becomes more pronounced, indicating a stronger infrared deformation. The parameter $\lambda$ specifies the radial location of the transition, effectively shifting the position of the step along the bulk direction and thereby setting the characteristic energy scale in the dual field theory. The parameter $\kappa$ governs the steepness of the transition: small values of $\kappa$ correspond to smooth, gradual profiles, while larger values produce a sharper and more localized transition, approaching a step-like behavior. In all cases, the dilaton interpolates between two constant asymptotic regimes, with the transition region becoming increasingly localized as $\kappa$ increases, emphasizing the role of scale separation in the holographic description.
\section{Schwinger effect with step Dilaton background}\label{sec:Dilaton}
The holographic description of the Schwinger effect is rooted in the correspondence between vacuum pair production in the boundary field theory and the dynamics of fundamental strings in the bulk geometry \cite{adscft}. In this setting, a virtual quark--antiquark pair is represented by the endpoints of an open string attached to a probe D3-brane, while the corresponding string worldsheet encodes the structure of the effective potential barrier governing the tunneling process \cite{Semenoff2011}. The pair production rate is determined by the competition between the external electric field, which tends to stretch the string and pull its endpoints apart, and the confining properties of the bulk background, which resist this separation. Accordingly, the classical string configuration extending from the probe brane into the bulk plays a decisive role in the holographic analysis. It provides a geometric realization of the quark--antiquark system and allows one to extract the total potential relevant for vacuum decay. Within this framework, the computation of the Coulomb potential naturally follows from the evaluation of the Wilson loop, which we consider in the following.
 
Our objective is to determine the Coulomb potential via a holographic evaluation of the rectangular Wilson loop on the probe D3-brane. This is implemented by computing the classical action of an open string whose endpoints are anchored on the probe D3-brane \cite{Semenoff2011}. Such a configuration constitutes a direct holographic realization of the Wilson loop prescription, analogous to the familiar circular loop construction, while offering a technically simpler yet equally effective setup for extracting the quark--antiquark potential in the given background.\\
Consider the background metric \eqref{metric} as $G_{\nu\mu}$ and  $(\nu,\mu)= (t, x, r)$ with components ,  
\begin{equation*}
 G_{tt}(r) =-f(r)\,\frac{r^2}{L^2} , \quad G_{xx_i}(r)=\frac{r^2}{L^2} , \quad G_{rr}(r) = \frac{L^2}{r^2\,f(r)}.
\end{equation*}
We choose the string worldsheet coordinates as \(\sigma^a = (\tau, \sigma)\) and impose the static gauge,
\begin{equation}\label{gauge}
t = \tau, \quad\quad x = \sigma.
\end{equation}
The radial coordinate of the classical string profile depends only on \(\sigma\),
\begin{equation}\label{eta}
r = r(\sigma).
\end{equation}
The Nambu-Goto Lagrangian density is given by,
\begin{equation}\label{Lresult}
\mathcal{L} =e^{\phi(r)/2}\, \sqrt{-G_{tt} (G_{xx} + G_{rr} r'^2)}=e^{\phi(r)/2}\,\sqrt{r'^2+\,f(r)\,\frac{r^4}{L^4}},
\end{equation}
where \(r' = \partial_\sigma r\).
Since the Lagrangian does not explicitly depend on \(\sigma\), the Hamiltonian density is conserved, yielding,
\begin{equation}\label{Lconstant}
\mathcal{L} - r' \frac{\partial \mathcal{L}}{\partial r'} = e^{\phi(r)/2}\,\sqrt{r'^2+\,f(r)\,\frac{r^4}{L^4}} -e^{\phi(r)/2}\, \frac{r'^2}{\sqrt{r'^2+\,f(r)\,\frac{r^4}{L^4}                                                                                                                                                                                         }} = \text{constant}.
\end{equation}
Applying the boundary condition at the turning point,
\begin{equation}\label{Lbound}
r = r_c, \quad r' = 0,
\end{equation}
fixes the constant to be,
\begin{equation}\label{consterm}
\text{constant} \equiv \frac{r^2_c}{L^2}\,e^{\phi(r_c)/2}\,  \sqrt{f(r_c)},
\end{equation}
therefore \eqref{Lconstant} leads to,
\begin{equation}\label{rprim}
r'^2=\frac{r^4}{L^4}\,f(r)\,\Big{(} \frac{e^{\phi(r)}\,f(r)\,r^4}{e^{\phi(r_c)}\,f(r_c)\,r_c^4}-1 \Big{)},
\end{equation}
also the differential equation for the profile is obtained as,
\begin{equation}\label{detadx}
\frac{dr}{dx} =\frac{r^2}{L^2}\, \sqrt{f(r)}\, \sqrt{\frac{e^{\phi(r)}\,f(r)\,r^4-e^{\phi(r_c)}\,f(r_c)\,r_c^4}{e^{\phi(r_c)}\,f(r_c)\,r_c^4}},
\end{equation}
or equivalently,
\begin{equation}\label{dxtodr}
dx =dr\,\frac{L^2}{r^2\, \sqrt{f(r)}}\,\sqrt{\frac{e^{\phi(r_c)}\,f(r_c)\,r_c^4}{e^{\phi(r)}\,f(r)\,r^4-e^{\phi(r_c)}\,f(r_c)\,r_c^4}}.
\end{equation}
Integrating, the half-separation between the pair endpoints is,
\begin{equation}
\int_0^{\frac{x}{2}} dx = L^2\,r^2_c e^{\phi(r_c)/2}\,\sqrt{f(r_c)} \int_{r_c}^{r_0}\frac{dr}{r^2\,\sqrt{f(r)\Big{(} e^{\phi(r)}\,f(r)\,r^4-e^{\phi(r_c)}\,f(r_c)\,r_c^4    \Big{)}}},
\end{equation}
that leads to,
\begin{equation}\label{xvfc}
x = 2 L^2\,r^2_c e^{\phi(r_c)/2}\,\sqrt{f(r_c)} \int_{r_c}^{r_0}\frac{dr}{r^2\,\sqrt{f(r)\Big{(} e^{\phi(r)}\,f(r)\,r^4-e^{\phi(r_c)}\,f(r_c)\,r_c^4    \Big{)}}}.
\end{equation}
Plugging \eqref{rprim} into \eqref{Lresult} results in,
\begin{equation}\label{lagrangyfin}
\mathcal{L}=\frac{e^{\phi(r)/2}}{L^2}\,\sqrt{\frac{e^{\phi(r)}\,f^2(r)\,r^8}{e^{\phi(r_c)}\,f(r_c)\,r^4_c}}.
\end{equation}
By using \eqref{lagrangyfin} and \eqref{dxtodr} the total potential energy combining the Coulomb potential and the static energy is given by,
\begin{equation}\label{V}
V_{CP+SE} =\, 2\, T_F \int^{\frac{x}{2}}_{0} dr\, \mathcal{L} = 2\,T_F \int_{r_c}^{r_0} dr \frac{e^{\phi(r)}\sqrt{f(r)\,r^4}}{\sqrt{e^{\phi(r)}\,f(r)\,r^4-e^{\phi(r_c)}\,f(r_c)\,r_c^4  }}.
\end{equation}
The critical electric field without magnetic field reads,
\begin{equation}\label{Ecr}
E_{cr} = T_F\,e^{\phi(r_0)/2} \sqrt{-G_{tt}(r_0) G_{xx}(r_0)} = T_F\,e^{\phi(r_0)/2} \frac{r^2_0}{L^2} \sqrt{f(r_0)}.
\end{equation}
The inclusion of an external magnetic field affects the holographic Schwinger mechanism through its appearance in the string effective action on the probe brane. In the probe brane setup, the endpoints of the fundamental string are coupled to the worldvolume gauge field living on the D3-brane. Consequently, the field strength \(F_{\nu\mu}\), containing both electric and magnetic components, enters the dynamics via the Dirac--Born--Infeld (DBI) action. Upon reduction to the string worldsheet, this contribution modifies the effective Nambu--Goto action and, in turn, alters the induced metric experienced by the string. The magnetic field therefore deforms the potential energy profile of the quark--antiquark pair already at the level of the worldsheet description, prior to any explicit analysis of the potential barrier or tunneling process. This framework has been systematically formulated in \cite{satoyo1303,Bolo1210,Koji1403}, where the combined effects of electric and magnetic fields are consistently incorporated into the holographic string configuration.\\
 When a magnetic field is present, the critical electric field is modified as,
 \begin{eqnarray}\label{EcrB}
E_{cr} &=& T_F\,e^{\phi(r_0)/2}\, \sqrt{-G_{tt}(r_0) G_{xx}(r_0)} \sqrt{1 + \frac{B_\perp^2}{T_F^2\,e^{\phi(r_0)}\,  G^2_{xx}(r_0) + B_\parallel^2}} \nonumber \\
&=& T_F\,\,e^{\phi(r_0)/2} \frac{r^2_0}{L^2}\, \sqrt{f(r_0)} \sqrt{1 + \frac{B_\perp^2}{T_F^2 \,e^{\phi(r_0)} \frac{r^4_0}{L^4} + B_\parallel^2}}, \quad B \neq 0,
\end{eqnarray}
\(B_\perp\) and \(B_\parallel\) denote the components of the magnetic field perpendicular and parallel to the electric field, respectively.
Defining the dimensionless ratio,
\begin{equation}\label{alpha}
\beta = \frac{E}{E_{cr}} \Rightarrow E = \beta\, E_{cr},
\end{equation}
the total potential including the effect of the electric field is,
\begin{equation}\label{vtot}
V_{tot} = V_{CP+SE} - E x=V_{CP+SE} - \beta\, E_{cr} x.
\end{equation}
Inserting the explicit expressions from \eqref{V}, \eqref{EcrB} and \eqref{xvfc} , one obtains,
\begin{eqnarray}\label{vtotfinal}
V_{tot} &=&2\, T_F \int_{r_c}^{r_0} dr   \Bigg{(}\frac{e^{\phi(r)}\sqrt{f(r)\,r^4}}{\sqrt{e^{\phi(r)}\,f(r)\,r^4-e^{\phi(r_c)}\,f(r_c)\,r_c^4  }}\nonumber\\
 &-&\beta\,e^{\phi(r_0)/2} r^2_0\, \sqrt{f(r_0)}\,e^{\phi(r_c)/2} r^2_c\, \sqrt{f(r_c)} \sqrt{1 + \frac{B_\perp^2}{T_F^2 \,e^{\phi(r_0)} \frac{r^4_0}{L^4} + B_\parallel^2}}\nonumber\\
& \times &\frac{1}{r^2\,\sqrt{f(r)\Big{(} e^{\phi(r)}\,f(r)\,r^4-e^{\phi(r_c)}\,f(r_c)\,r_c^4    \Big{)}}}     \Bigg{)}
\end{eqnarray}
Introducing the dimensionless variables,
\begin{equation}\label{aby}
a = \frac{r_c}{r_0}, \quad b = \frac{r_h}{r_0}, \quad y = \frac{r}{r_c} = \frac{r}{a r_0},
\end{equation}
and applying them to \eqref{xvfc} and \eqref{vtotfinal}, they take the form,

\begin{equation}\label{xy}
x =\frac{2 L^2}{a\,r_0} e^{\phi(y_c)/2}\,\sqrt{1-\frac{b^4}{a^4}}\,\int_{1}^{1/a} \frac{dy}{\sqrt{(y^4-\frac{b^4}{a^4})
\Big{(}e^{\phi(y)}\,(y^4-\frac{b^4}{a^4})-e^{\phi(y_c)}\,(1-\frac{b^4}{a^4})\Big{)}}},
\end{equation}
and,
\begin{eqnarray}\label{vtotfinalaby}
V_{tot}&=&2\,T_F\,a\,r_0\,\int_{1}^{1/a} dy\, \Bigg{(}  \frac{e^{\phi(y)}\,\sqrt{y^4-\frac{b^4}{a^4}}}{\sqrt{e^{\phi(y)}\,(y^4-\frac{b^4}{a^4})-e^{\phi(y_c)}\,(1-\frac{b^4}{a^4})}} \nonumber\\
&-&\beta\, e^{\phi(y_0)/2}\, \sqrt{1-b^4}\,e^{\phi(y_c)/2}\,\sqrt{1-\frac{b^4}{a^4}}  \sqrt{1+\frac{B_\perp^2}{\frac{T_F^2}{L^4} \, e^{\phi(y_0)}\,r_0^4\,+B_\parallel^2 }} \nonumber\\
&\times &  \frac{1}{a^2\, \sqrt{y^4-\frac{b^4}{a^4}}\,\sqrt{e^{\phi(y)}\,(y^4-\frac{b^4}{a^4})-e^{\phi(y_c)}\,(1-\frac{b^4}{a^4})}}   \Bigg{)}
\end{eqnarray}

Now, consider replacing a probe D3-brane at an intermediate
position $r = r_0$ rather than close to the boundary. The mass is the energy of a single string stretching between
the probe D3-brane at $r = r_0$ and the horizon,  then the mass becomes finite and
depends on $r_0$ as,
\begin{equation}\label{mass}
m=T_F\,e^{\phi(r_0)/2}\,\int^{r_0}_{r_h}\, dr \sqrt{det\, g_{tr} }= T_F\,e^{\phi(r_0)/2}\,\int^{r_0}_{r_h}\, dr =T_F\,e^{\phi(r_0)/2}\,(r_0-r_h),
\end{equation}
where we used the induced metric for the string as,
\begin{equation}
g_{tr}=diag(G_{tt}(r), G_{rr}(r)).
\end{equation}
\subsection{Vacuum instability in the absence of  Magnetic Field}
Considering the total potantial \eqref{vtotfinalaby} versus $x$ we study the schwinger effect in the absence of  magnetic Field.
\subsubsection{Subcritical Electric Field Regime $\beta<1 $}
\begin{figure}[h!]
\begin{minipage}[c]{1\textwidth}
\tiny{(a)}\includegraphics[width=8cm,height=5cm,clip]{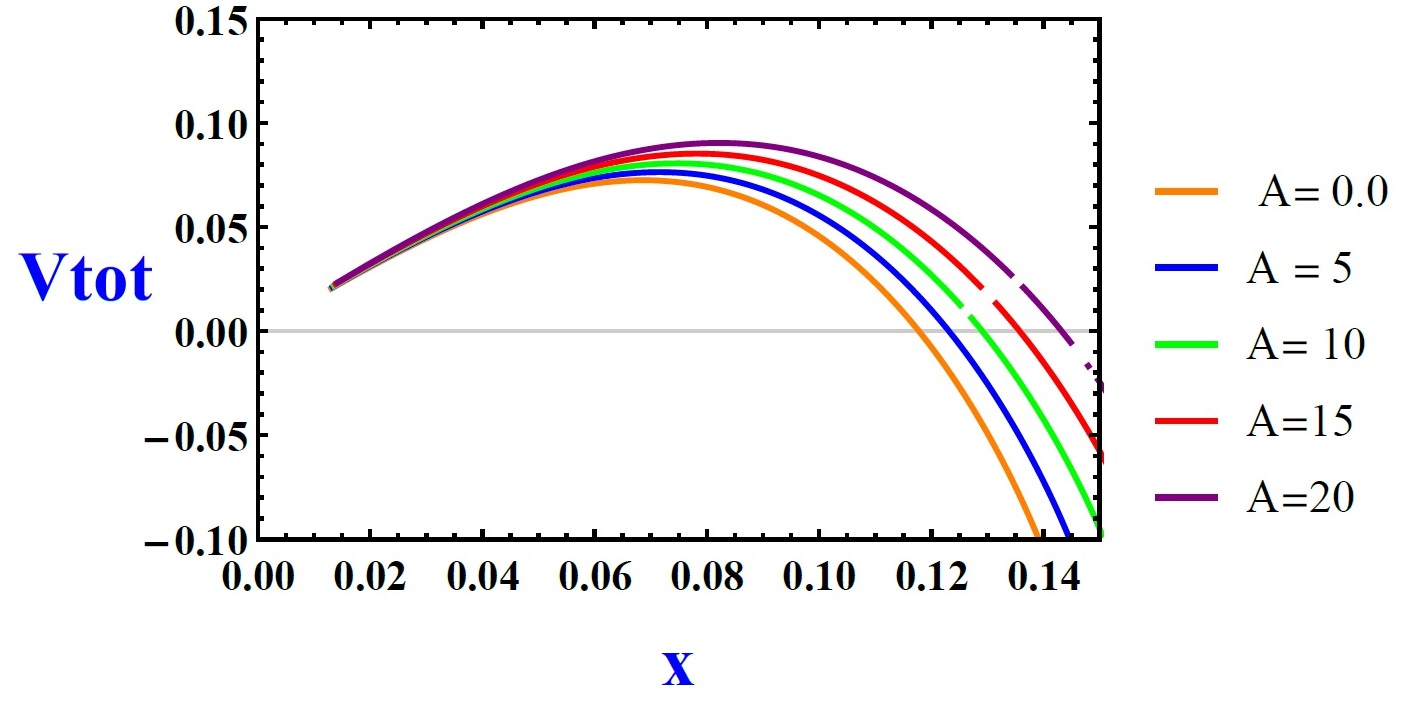}
\hspace{0.1cm}
\tiny{(b)}\includegraphics[width=8cm,height=5cm,clip]{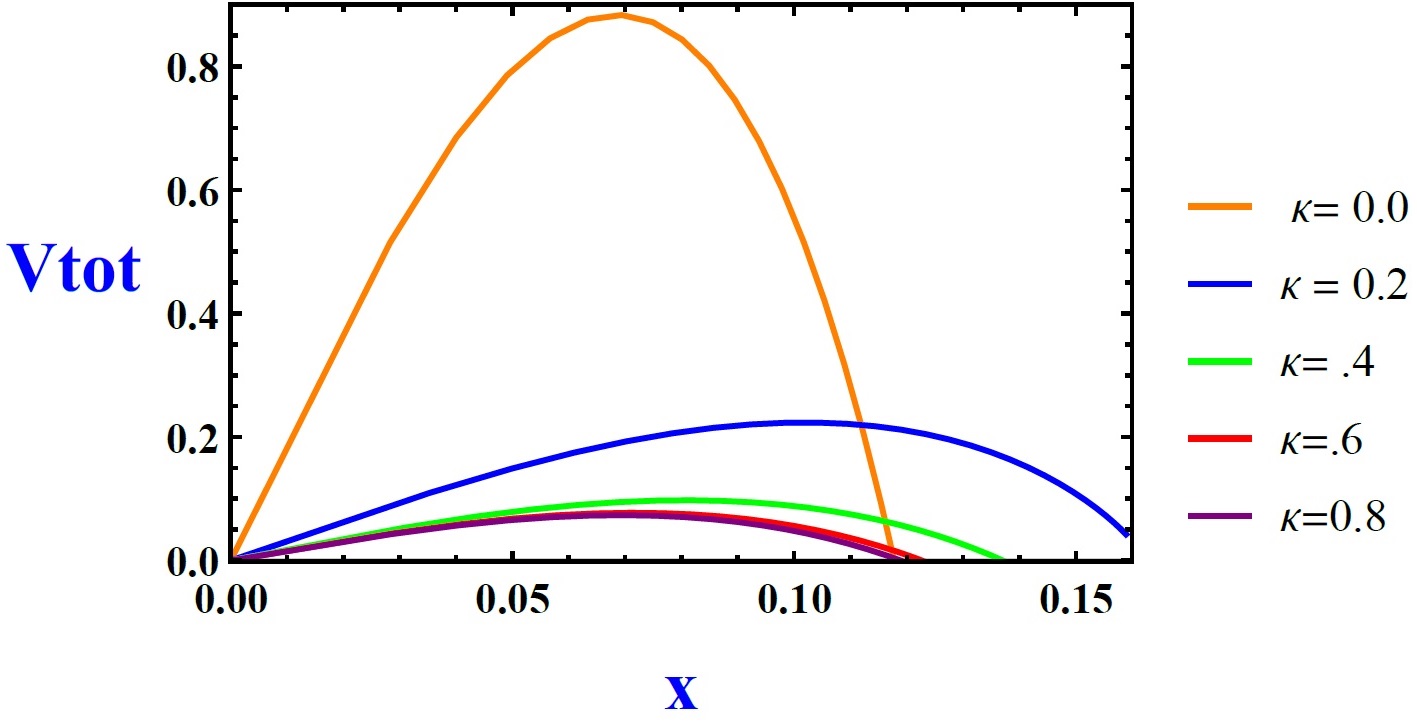}
\end{minipage}
\caption{The total potential in the $\beta<1$ regime in the absence of a magnetic field: (a) for different values of $A$, and (b) for different values of $\kappa$.}
\label{3,kappa3}
\end{figure}
\textbf{Fig.~\ref{3,kappa3} }  presents the total potential $V_{\text{tot}}$ in the subcritical regime ($\beta<1$), where a potential barrier is present. Panel (a) illustrates the effect of varying $A$, while panel (b) displays the dependence on the parameter $\kappa$.
 
In panel (a), the curves show that increasing $A$ systematically enhances the potential barrier. The curves corresponding to larger $A$ lie above the others, with a higher maximum and a steeper rise near the peak. This behavior follows directly from Fig.~\ref{fig:path_independence}, where increasing $A$ raises the infrared (IR) value of the dilaton. Through the factor $e^{\phi(r)/2}$ in the Nambu--Goto action, this leads to a larger effective string tension in the IR region. Consequently, the string configuration encounters stronger resistance when extending into the bulk, resulting in a higher and sharper barrier. Physically, this corresponds to stronger confinement and a reduced probability of pair production.
 In panel (b), the effect of $\kappa$ is qualitatively different. The curves indicate that increasing $\kappa$ lowers the maximum of the potential barrier while extending it over a wider range of separation. In other words, the barrier becomes less pronounced but more spread out along the horizontal axis. Referring again to Fig.~\ref{fig:path_independence}, larger values of $\kappa$ make the dilaton transition sharper but confine it to a narrower region around $r\sim\lambda$. As a result, the enhancement of the string tension due to the IR region becomes more localized. Since the string worldsheet probes this region only over a limited interval in $r$, the integrated contribution of the strongly coupled region to the total energy is reduced. This leads to a suppression of the barrier height, while the overall structure of the potential becomes broader.
 
In contrast to the standard soft-wall model, where the dilaton typically grows smoothly and monotonically in the IR, the present profile introduces a controlled and localized transition. This allows one to disentangle the magnitude of IR effects (through $A$) from their localization (through $\kappa$), leading to qualitatively different modifications of the potential barrier.

\subsubsection{Critical Electric Field Regime $\beta=1$}

\begin{figure}[h!]
\begin{minipage}[c]{1\textwidth}
\tiny{(a)}\includegraphics[width=8cm,height=5cm,clip]{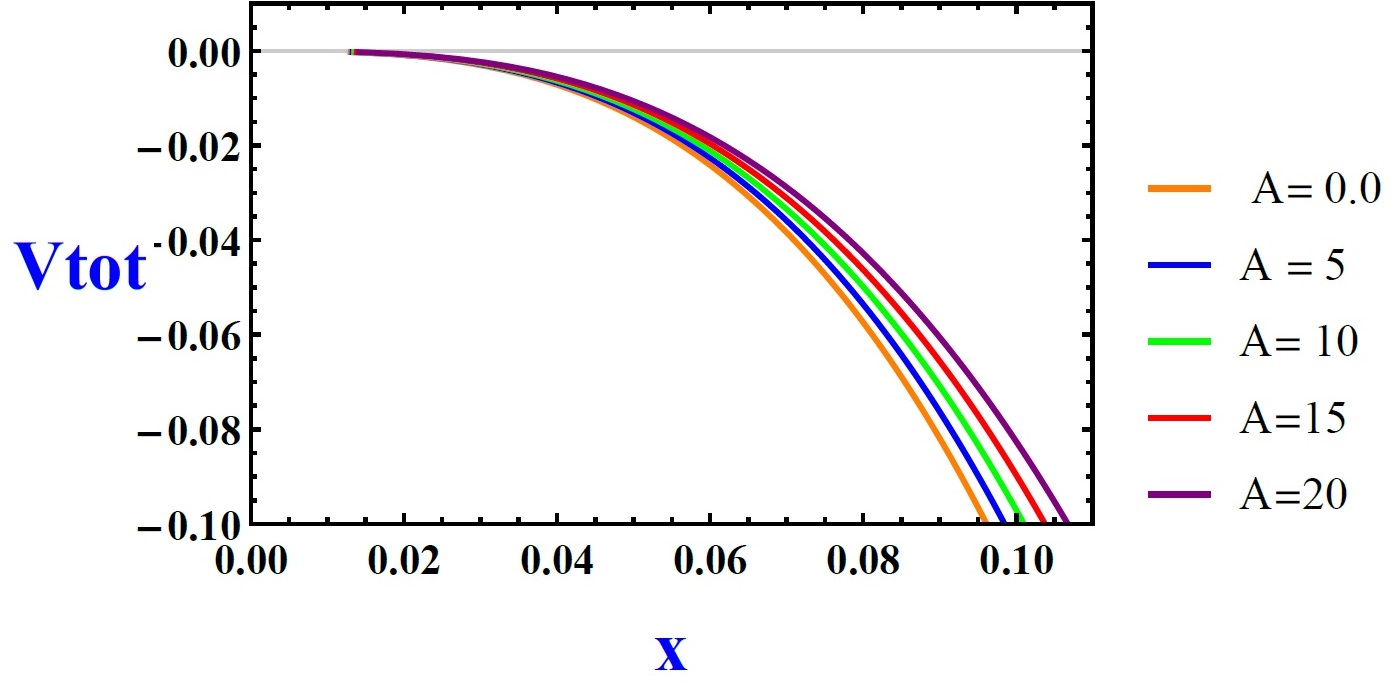}
\hspace{0.2cm}
\tiny{(b)}\includegraphics[width=8cm,height=5cm,clip]{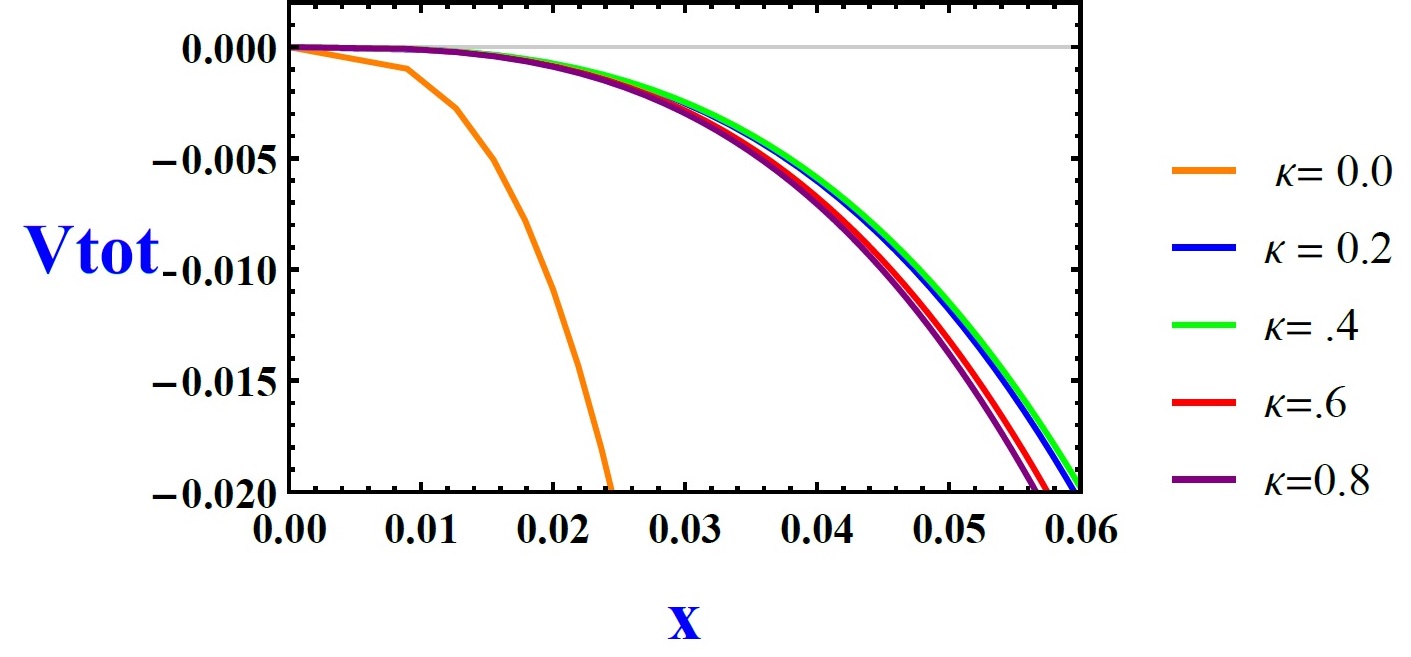}
\end{minipage}
\caption{The total potential in the $\beta=1$ regime in the absence of a magnetic field: (a) for different values of $A$, and (b) for different values of $\kappa$.}
\label{2,kappa2}
\end{figure}
\textbf{Fig.~\ref{2,kappa2}} illustrates the total potential $V_{\text{tot}}$ at the critical point ($\beta=1$), where the potential barrier disappears and the system reaches the threshold of instability. Panel (a) corresponds to varying $A$, while panel (b) shows the dependence on $\kappa$.
 
In panel (a), increasing $A$ shifts the potential upward across the entire separation range. This behavior is directly tied to Fig.~\ref{fig:path_independence}, where larger values of $A$ enhance the IR magnitude of the dilaton. Consequently, the effective string tension is strengthened, and the potential grows more prominently with separation even at the critical point. This reflects the persistence of strong IR effects at the onset of instability.
 In panel (b), the dependence on $\kappa$ is non-monotonic. As $\kappa$ increases from small values up to intermediate values of $\kappa$, the potential shifts upward and attains its maximum profile, indicating an enhancement of the effective interaction. However, for larger $\kappa$, the curves move downward again. This behavior can be understood by referring to Fig.~\ref{fig:path_independence}: increasing $\kappa$ sharpens the dilaton transition around $r\sim\lambda$. For intermediate values (such as $\kappa=0.4$), the string worldsheet still samples both UV and IR regions effectively, leading to a maximal contribution from the IR-enhanced coupling. In contrast, for larger $\kappa$, the transition becomes highly localized, reducing the extent to which the string probes the IR region, and thereby weakening the effective string tension.
 
At $\beta=1$, the system lies precisely at the boundary between stable and unstable configurations. Compared to the soft-wall scenario, where a smooth quadratic dilaton profile enforces a monotonic modification of the potential, the present model exhibits a more intricate dependence on $\kappa$, reflecting the interplay between the magnitude and localization of IR effects.

\subsubsection{Supercritical Electric Field Regime $\beta>1 $}

\begin{figure}[h!]
\begin{minipage}[c]{1\textwidth}
\tiny{(a)}\includegraphics[width=8cm,height=5cm,clip]{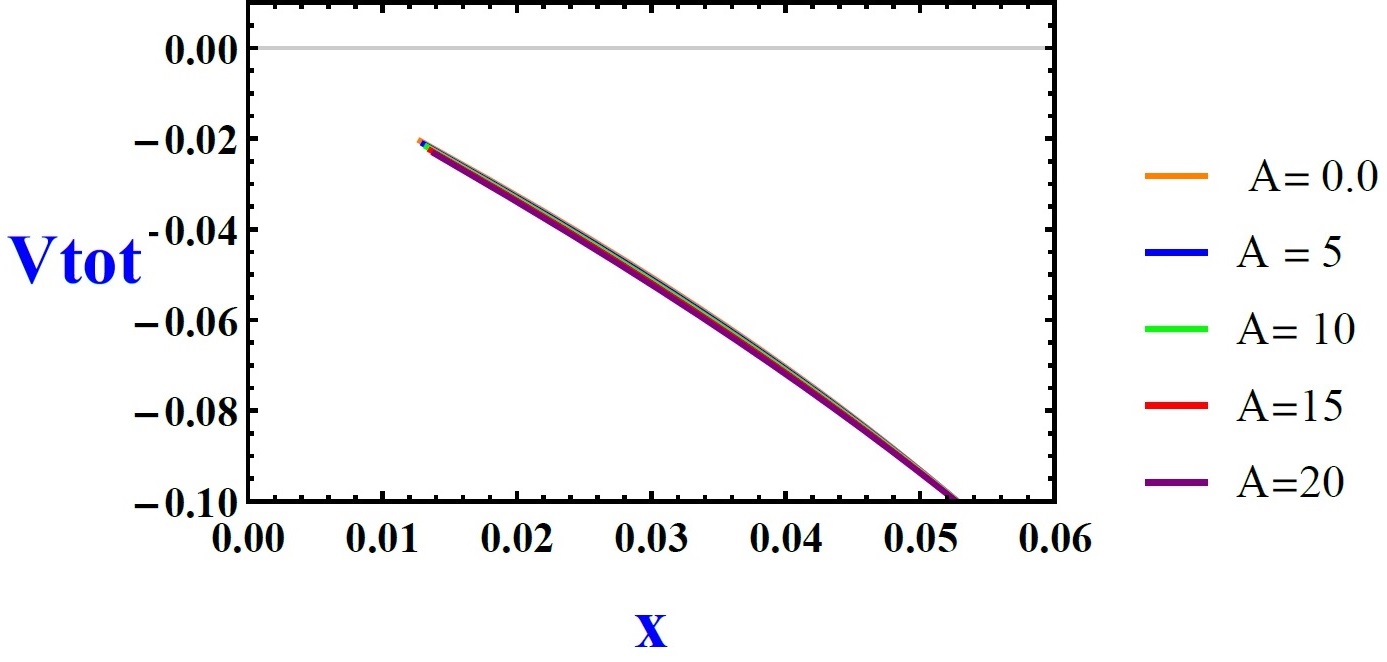}
\hspace{0.2cm}
\tiny{(b)}\includegraphics[width=8cm,height=5cm,clip]{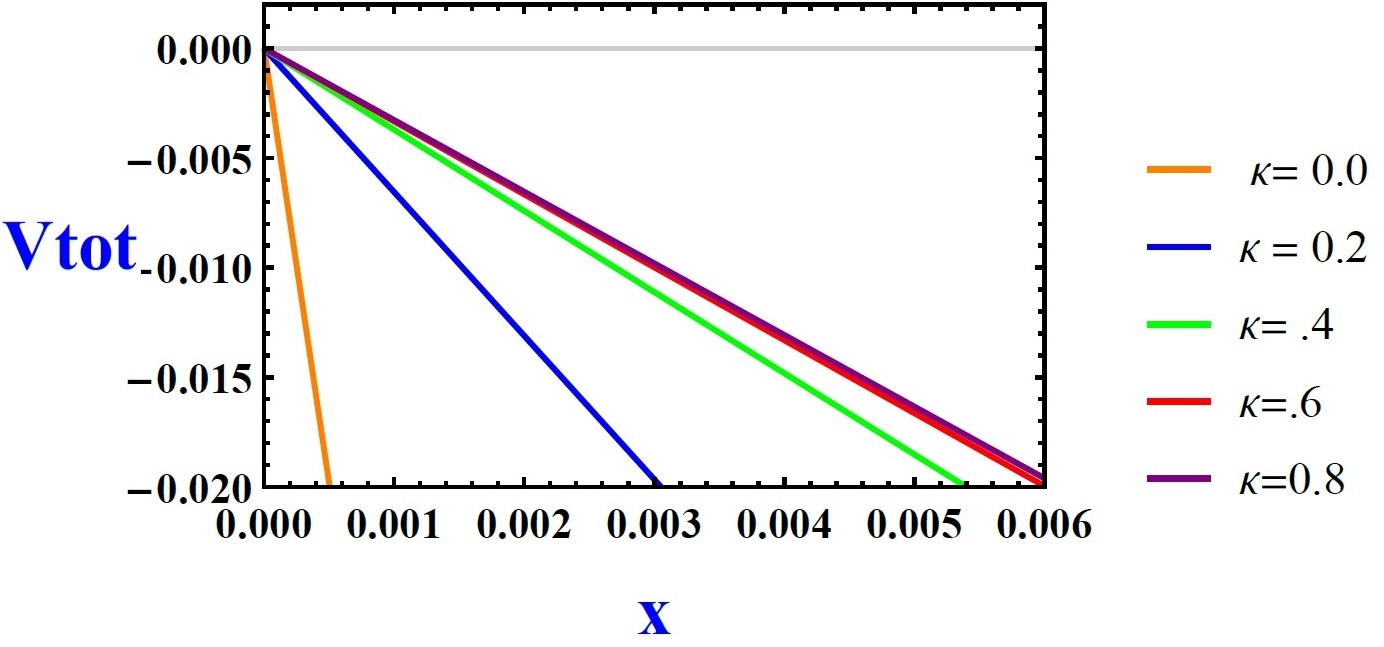}
\end{minipage}
\caption{The total potential in the $\beta>1$ regime in the absence of a magnetic field: (a) for different values of $A$, and (b) for different values of $\kappa$.}
\label{1, kappa1}
\end{figure}
\textbf{Fig.~\ref{1, kappa1}} displays the total potential $V_{\text{tot}}$ in the supercritical regime ($\beta>1$), where the potential barrier is absent and the system is fully unstable. Panel (a) corresponds to varying $A$, while panel (b) shows the dependence on $\kappa$.
 
 In panel (a), all curves corresponding to different values of $A$ collapse onto each other and become practically indistinguishable. This indicates that, in the supercritical regime, the effect of $A$ on the total potential is effectively washed out. Although increasing $A$ enhances the IR value of the dilaton, as seen in Fig.~\ref{fig:path_independence}, this contribution becomes subdominant once the external electric field exceeds the critical value. In this regime, the linear term associated with the electric field dominates the potential, and the sensitivity to the IR deformation is significantly reduced. As a result, variations in $A$, which primarily modify the strength of the IR physics, no longer produce visible differences in the potential.
 In panel (b), the dependence on $\kappa$ becomes monotonic, in contrast to the behavior observed at the critical point. The curves show that increasing $\kappa$ shifts the potential upward and reduces its slope. Referring to Fig.~\ref{fig:path_independence}, larger values of $\kappa$ localize the dilaton transition more sharply around $r\sim\lambda$, thereby restricting the extent of the IR-enhanced region. In the supercritical regime, where the barrier has already disappeared, the string configuration is more sensitive to the overall reduction of the IR contribution rather than to its detailed distribution. Consequently, increasing $\kappa$ leads to a continuous weakening of the effective string tension, reflected in the monotonic decrease in potential magnitude.
 
The difference between this behavior and the critical case can be understood as follows: at $\beta=1$, the system is sensitive to a balance between competing effects, which allows a non-monotonic dependence on $\kappa$ to emerge. In contrast, for $\beta>1$, the dominance of the electric field removes this balance, and the response of the system becomes simpler and monotonic. Compared to the soft-wall model, where the dilaton profile induces a smooth and uniform modification, the present setup shows that once the system enters the unstable regime, only the overall suppression of IR effects remains relevant, while finer structural details of the dilaton profile play a diminished role.

\subsection{Vacuum instability in the presence of  Magnetic Field}
In this subsection, we consider the total potential \eqref{vtotfinalaby} in the presence of an external magnetic field and examine its impact on the Schwinger effect.

\subsubsection{Subcritical Electric Field Regime $\beta<1 $ in the presence of $B\neq 0$}
\begin{figure}[h!]
\begin{center}$
\begin{array}{cccc}
\includegraphics[width=10cm]{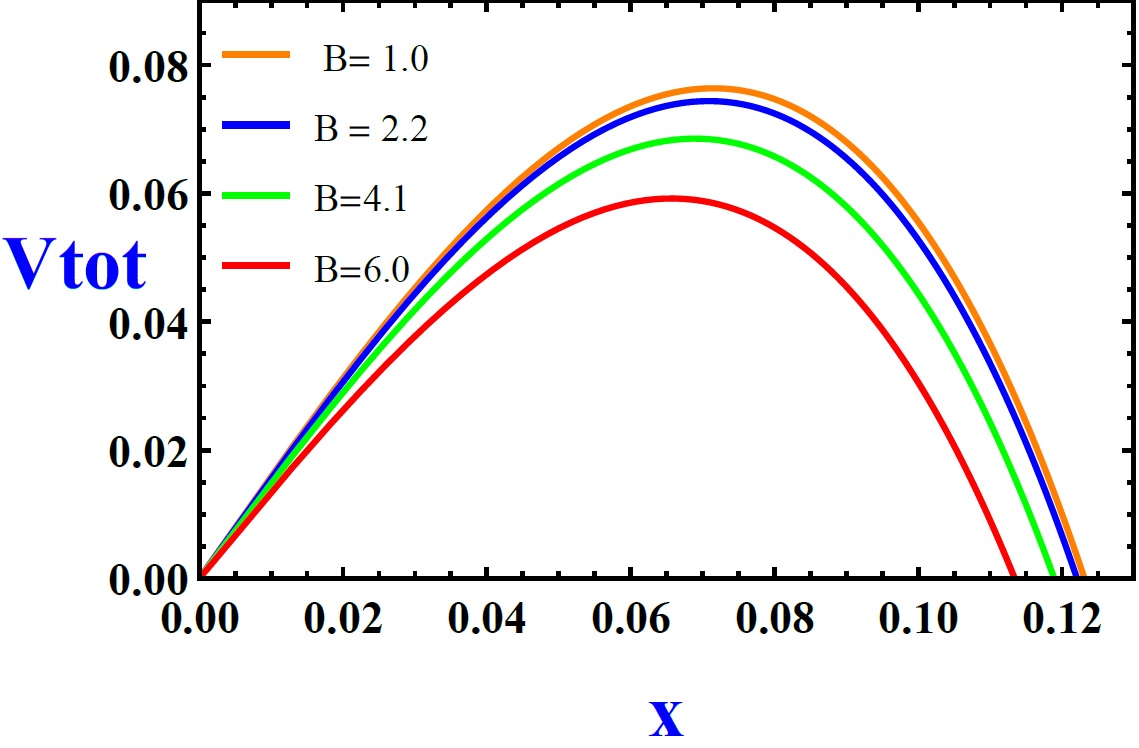}
\end{array}$
\end{center}
\caption{The total potential in the $\beta<1$ regime in the presence of a magnetic field for different values of the magnetic field. The parameters are fixed to $r_0=5$, $b=8$, $A=5$, $\lambda=0.2$, $L=1$, and $\kappa=1$.}
\label{Beta0.9Bchanges}
\end{figure}
\textbf{Fig.~\ref{Beta0.9Bchanges}} presents the total potential $V_{\text{tot}}$ in the subcritical regime ($\beta<1$) in the presence of a magnetic field, for different values of $B$.
 
The curves show that increasing $B$ reduces the height of the potential barrier and shifts it downward. The maximum of the potential becomes lower, indicating that the tunneling barrier is progressively suppressed as the magnetic field increases.
 This behavior indicates that, in the subcritical regime, the magnetic field facilitates pair production by weakening the barrier that opposes separation. Although the system is still below the critical electric field and a barrier exists, increasing $B$ reduces the energy required for tunneling.
 From a physical point of view, this reflects a modification in the balance between the electric contribution and the string energy in the total potential. The magnetic field effectively diminishes the contribution responsible for maintaining the barrier, thereby driving the system closer to the onset of instability.
 
In comparison with Fig.~\ref{fig:path_independence}, this effect is qualitatively different from increasing $A$, which enhances the IR contribution and typically strengthens the barrier. Here, the magnetic field acts as an external parameter that suppresses the barrier without altering the intrinsic dilaton profile. 
\subsubsection{Critical Electric Field Regime $\beta=1$ in the presence of $B\neq 0$}

\begin{figure}[h!]
\begin{center}$
\begin{array}{cccc}
\includegraphics[width=10cm]{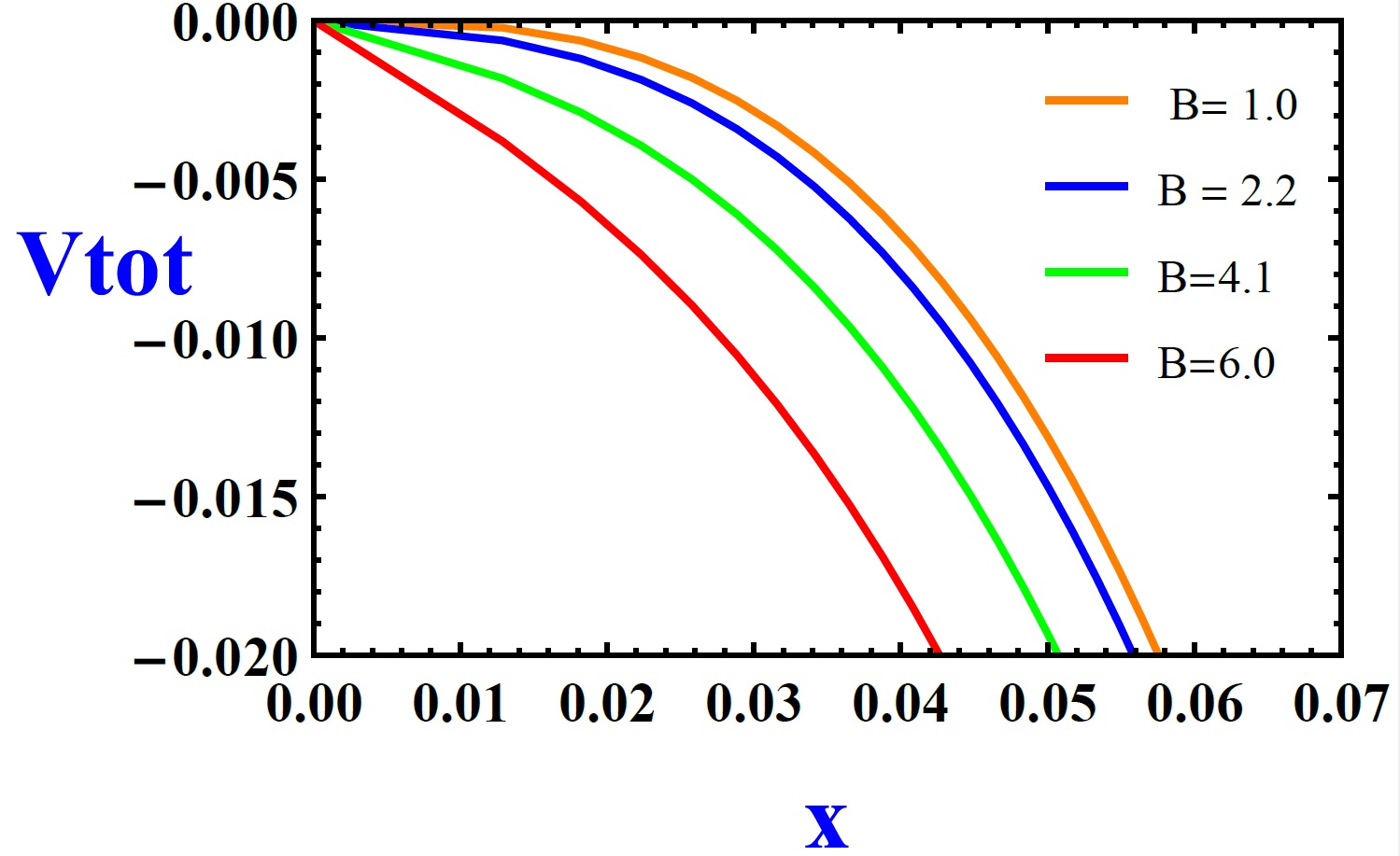}
\end{array}$
\end{center}
\caption{The total potential in the $\beta=1$ regime in the presence of a magnetic field for different values of the magnetic field. The parameters are fixed to $r_0=5$, $b=8$, $A=5$, $\lambda=0.2$, $L=1$, and $\kappa=1$.}
\label{Beta1.0Bchanges}
\end{figure}
\textbf{Fig.~\ref{Beta1.0Bchanges}} illustrates the total potential $V_{\text{tot}}$ at the critical point ($\beta=1$) in the presence of a magnetic field, for different values of $B$.
 
The curves show that increasing $B$ shifts the potential downward over the entire range of separation, while at the same time making the curves steeper. The configurations corresponding to larger $B$ lie below the others but increase more sharply with separation. No barrier is formed at this critical value of $\beta$.
 This behavior reflects the interplay between the electric contribution and the string energy in the total potential. At $\beta=1$, these two contributions are balanced in such a way that the barrier disappears. The magnetic field modifies this balance by reducing the overall energy scale of the configuration, while the dependence on separation remains governed by the string dynamics. As a result, the potential is shifted downward, but its variation with separation becomes more pronounced.
 
In comparison with Fig.~\ref{fig:path_independence}, this effect is qualitatively different from varying the dilaton parameters, which directly control the IR structure of the background. Here, the magnetic field acts as an external parameter that alters the resulting potential without modifying the dilaton profile itself.
 
At the critical point, the system is highly sensitive to external deformations, and therefore even moderate values of $B$ can significantly reshape the potential. 
\subsubsection{Supercritical Electric Field Regime $\beta>1 $ in the presence of $B\neq 0$}
\begin{figure}[h!]
\begin{center}$
\begin{array}{cccc}
\includegraphics[width=10cm]{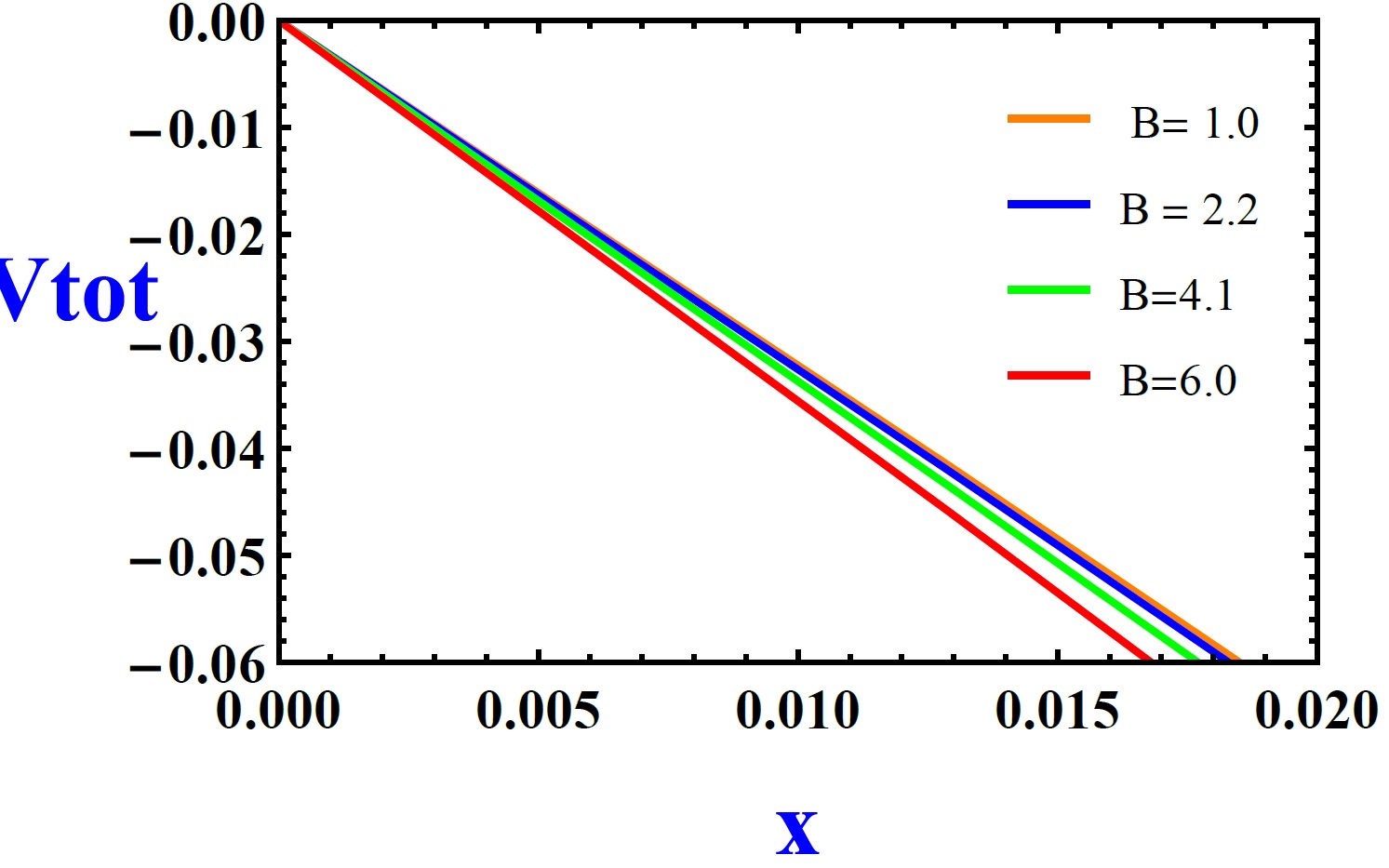}
\end{array}$
\end{center}
\caption{The total potential in the $\beta>1$ regime in the presence of a magnetic field for different values of the magnetic field. The parameters are fixed to $r_0=5$, $b=8$, $A=5$, $\lambda=0.2$, $L=1$, and $\kappa=1$.}
\label{Beta1.2Bchanges}
\end{figure}
\textbf{Fig.~\ref{Beta1.2Bchanges}} shows the total potential $V_{\text{tot}}$ in the supercritical regime ($\beta>1$) in the presence of a magnetic field, for different values of $B$.
 
The curves indicate that increasing $B$ shifts the potential downward over the entire range of separation. Configurations with larger $B$ lie systematically below those with smaller magnetic field. As expected in the supercritical regime, no potential barrier is present.
 This behavior reflects the dominance of the electric field together with the modifying role of the magnetic field. Since the system is already beyond the critical point, the barrier has disappeared and the dynamics are governed by the competition between the electric contribution and the string energy. The magnetic field reduces the overall energy scale of the configuration, leading to the observed downward shift of the potential.
 
Compared to the critical case ($\beta=1$), the response becomes more uniform due to the absence of any residual balance between competing effects. The system no longer exhibits threshold sensitivity, and the influence of $B$ appears as a smooth and monotonic deformation of the potential.
 
In relation to Fig.~\ref{fig:path_independence}, the effect of the magnetic field remains distinct from changes in the dilaton parameters. While the latter modify the IR structure of the background, the magnetic field acts externally and reshapes the resulting potential without altering the underlying profile. 






 \section{Summary and Outlook}\label{sec:summary}
 
In this work, we have investigated the holographic Schwinger effect in a background characterized by a step dilaton profile, which provides a sharp realization of confinement through a sudden transition between ultraviolet and infrared regimes. Within this framework, we constructed the holographic setup and evaluated the total potential between a quark--antiquark pair from the classical configuration of a fundamental string, allowing us to probe the vacuum instability associated with pair production.
 
In the absence of a magnetic field, our analysis shows that the step dilaton profile leads to a significant modification of the potential barrier. As the external electric field increases, the barrier is progressively reduced and eventually vanishes at a critical value, signaling the onset of catastrophic vacuum decay. Compared to smooth soft-wall models, the presence of the step dilaton induces a noticeably sharper suppression of both the height and width of the barrier. Furthermore, the properties of the potential are highly sensitive to the parameters of the dilaton profile, indicating that the position and strength of the step play a crucial role in controlling the pair production process.
 
When an external magnetic field is introduced, the structure of the potential becomes considerably richer. We find that the magnetic field modifies the barrier in a nontrivial manner, leading to a clear shift in the critical electric field. Depending on its magnitude and orientation relative to the electric field, the magnetic field can either enhance or suppress the Schwinger effect by altering the height and width of the potential barrier. In the step dilaton background, this interplay is amplified due to the sharp geometric transition, resulting in a stronger sensitivity of the system to external fields compared to smooth geometries.
 
A detailed comparison with soft-wall models shows that the qualitative behavior of the Schwinger effect is strongly dependent on the form of the dilaton profile. While smooth backgrounds lead to gradual and moderate modifications of the potential, the step dilaton produces more abrupt changes and a markedly enhanced response to variations in both the electric and magnetic fields. This distinction highlights the importance of the underlying holographic structure in determining non-perturbative phenomena such as vacuum pair production.
 
In summary, our results demonstrate that the step dilaton background provides a qualitatively distinct and physically rich framework for the holographic Schwinger effect. The sharp transition in the dilaton profile leads to an enhanced sensitivity of the potential barrier to external electromagnetic fields, resulting in more pronounced modifications of the critical electric field and the pair production process relative to conventional soft-wall models.

\section*{Acknowledgement}
This work is supported by the National Natural Science Foundation of China (Grant No. 12275067),
Science and Technology R$\&$D Program Joint Fund Project of Henan Province  (Grant No.225200810030),
Science and Technology Innovation Leading Talent Support Program of
Henan Province  (Grant No. 254200510039),
and National Key R$\&$D Program of China (Grant No. 2023YFA1606000).


\begin{thebibliography}{99}
\bibitem{adscft}
 J. M. Maldacena, {\it{``The Large N limit of superconformal field theories and
supergravity''}}, \emph{Adv. Theor. Math. Phys.} {\bf 2}  (1998) 231 [arXiv:hep-th/9711200]. 
\bibitem{Karch2006}
Andreas Karch, Emanuel Katz, Dam T. Son, Mikhail A. Stephanov, {\it{``Linear Confinement and AdS/QCD''}}, \emph{Phys. Rev. D} {\bf 74} (2006) 015005 [arXiv:hep-ph/0602229].
\bibitem{Cox2015}
Peter Cox, Tony Gherghetta, {\it{``A Soft-Wall Dilaton''}}, \emph{JHEP} {\bf 02} (2015) 006
[arXiv:1411.1732 [hep-ph]].
\bibitem{Li2013}
Danning Li, Mei Huang, {\it{``Dynamical holographic QCD model for glueball and light meson spectra''}},\emph{JHEP} {\bf 11} (2013) 088 [arXiv:1303.6929[hep-ph]].
\bibitem{Semenoff2011}
Gordon W. Semenoff, Konstantin Zarembo, {\it{``Holographic Schwinger Effect''}},\emph{Phys. Rev. Lett.} {\bf 107} (2011)  171601 [arXiv:1109.2920 [hep-th]].
\bibitem{Sato2013}
Yoshiki Sato, Kentaroh Yoshida, {\it{``Potential Analysis in Holographic Schwinger Effect''}},\emph{JHEP} {\bf 08}  (2013) 002 [arXiv:1304.7917 [hep-th]].
 \bibitem{Sato1309}
Yoshiki Sato, Kentaroh Yoshida, {\it{“Universal aspects of holographic Schwinger effect in general backgrounds''}},\emph{ JHEP} {\bf 12} (2013)  051 [arXiv:1309.4629  [hep-th]].
\bibitem{Hou2018}
Le Zhang, De-Fu Hou, Jian Li, {\it{``Holographic Schwinger Effect with Chemical Potential at Finite Temperature''}}, \emph{Eur. Phys. J. A} {\bf 54} 94 (2018).
\bibitem{Zhang2016}
Zi-Qiang Zhang, De-Fu Hou, Yan Wu, Gang Chen, {\it{``Holographic Schwinger Effect in a Confining D3-Brane Background with Chemical Potential''}}, \emph{Adv. High Energy Phys.} {\bf 2016}  (2016) 9258106  [arXiv:1604.00095 [hep-ph]].
\bibitem{Kazem1504}
Kazem Bitaghsir Fadafan, Fateme Saiedi, {\it{``On Holographic Non-relativistic Schwinger Effect''}},\emph{Eur. Phys. J. C} {\bf 75}  612 (2015) [arXiv:1504.02432 [hep-th]].
\bibitem{Zi-qiang2018}
Zi-qiang Zhang, {\it{``Potential analysis in holographic Schwinger effect in Einstein–Maxwell–Gauss–Bonnet gravity''}}, \emph{Nucl. Phys. B} {\bf 935} (2018) 377.
\bibitem{Lin2025}
Sheng Lin,  Xuan Liu,  Xun Chen,  Gen-Fa Zhang,  Jing Zhou, {\it{``Holographic Schwinger Effect in Flavor-Dependent Systems''}}, \emph{Phys. Rev. D} {\bf 111}  046005  (2025) [arXiv:2407.14828 [hep-ph]].
\bibitem{Wen2506}
Wen-Bin Chang, Defu Hou, {\it{``Schwinger Effect in a Twice Anisotropic Holographic Model''}} \emph{Chin.Phys. 50  } {\bf  5, 053103 } (2026) [arXiv:2506.16245 [hep-ph]].
\bibitem{sara2025}
Sara Tahery  {\it{``Holographic Schwinger effect with Translational Symmetry Breaking''}},[arXiv:2510.13707 [hep-th]].
\bibitem{zhu2021}
Zhou-Run Zhu, Yang-Kang Liu, De-Fu Hou, {\it{``Holographic Schwinger Effect in the Dynamical AdS/QCD Model''}}, [arXiv:2108.05148 [hep-ph]].
\bibitem{satoyo1303}
Yoshiki Sato, Kentaroh Yoshida, {\it{``Holographic description of the Schwinger effect in electric and magnetic fields''}},\emph{JHEP} {\bf 04}   111 (2013)   [arXiv:1303.0112 [hep-th]].
 \bibitem{Bolo1210}
S. Bolognesi, F. Kiefer, E. Rabinovici, {\it{``Comments on Critical Electric and Magnetic Fields from Holography''}},\emph{JHEP} {\bf 01} 174  (2013)  [arXiv:1210.4170 [hep-th]].
\bibitem{Koji1403}
Koji Hashimoto, Takashi Oka, Akihiko Sonoda, {\it{``Magnetic instability in AdS/CFT : Schwinger effect and Euler-Heisenberg Lagrangian of Supersymmetric QCD''}},\emph{JHEP} {\bf 06} 085 (2014) [arXiv:1403.6336 [hep-th]].
\bibitem{zhou1912}
Zhou-Run Zhu, De-Fu Hou, Xun Chen, {\it{``Potential analysis of holographic Schwinger effect in the magnetized background''}}, \emph{Eur. Phys. J. C} {\bf 80} 550  (2020)  
[arXiv:1912.05806 [hep-ph]].
\bibitem{Ding2020}
Yue Ding, Zi-qiang Zhang, {\it{``Holographic Schwinger Effect in a Soft-Wall Model''}},  \emph{Chin. Phys. C} {\bf 45} (2021) 1, 013111 [arXiv:2009.06179 [hep-th]].
















\end{thebibliography}
\end{document}